\documentclass[]{spie}  %>>> use for US letter paper
\usepackage{amsmath,amsfonts,amssymb}
\usepackage{graphicx}
\usepackage[colorlinks=true, allcolors=blue]{hyperref}
\usepackage{adjustbox}

\title{Can patient-specific acquisition protocol improve performance on defect detection task in myocardial perfusion SPECT?}

\author[a]{Nu Ri Choi}
\author[b]{Md Ashequr Rahman}
\author[b]{Zitong Yu}
\author[c]{Barry A. Siegel}
\author[b,c]{Abhinav K. Jha}

\affil[a]{Department of Computer Science \& Engineering, Washington University, St. Louis, MO, USA}
\affil[b]{Department of Biomedical Engineering, Washington University, St. Louis, MO, USA}
\affil[c]{Mallinckrodt Institute of Radiology, Washington University, St. Louis, MO, USA}
\begin{document} 
\maketitle

\begin{abstract}
Myocardial perfusion imaging using single-photon emission computed tomography (SPECT), or myocardial perfusion SPECT (MPS) is a widely used clinical imaging modality for the diagnosis of coronary artery disease. Current clinical protocols for acquiring and reconstructing MPS images are similar for most patients. However, for patients with outlier anatomical characteristics, such as large breasts, images acquired using conventional protocols are often sub-optimal in quality, leading to degraded diagnostic accuracy. Solutions to improve image quality for these patients outside of increased dose or total acquisition time remain challenging. Thus, there is an important need for new methodologies that can help improve the quality of the acquired images for such patients, in terms of the ability to detect myocardial perfusion defects. One approach to improving this performance is adapting the image acquisition protocol specific to each patient. Studies have shown that in MPS, different projection angles usually contain varying amounts of information for the detection task. However, current clinical protocols spend the same time at each projection angle. In this work, we evaluated whether an acquisition protocol that is optimized for each patient could improve performance on the task of defect detection on reconstructed images for patients with outlier anatomical characteristics. For this study, we first designed and implemented a personalized patient-specific protocol-optimization strategy, which we term precision SPECT (PRESPECT). This strategy integrates the theory of ideal observers with the constraints of tomographic reconstruction to optimize the acquisition time for each projection view, such that performance on the task of detecting myocardial perfusion defects is maximized. We performed a clinically realistic simulation study on patients with outlier anatomies on the task of detecting perfusion defects on various realizations of low-dose scans by an anthropomorphic channelized Hotelling observer. Our results show that using PRESPECT led to improved performance on the defect detection task for the considered patients. These results provide evidence that personalization of MPS acquisition protocol has the potential to improve defect detection performance on reconstructed images by anthropomorphic observers for patients with outlier anatomical characteristics. Thus, our findings motivate further research to design optimal patient-specific acquisition and reconstruction protocols for MPS, as well as developing similar approaches for other medical imaging modalities. 
\end{abstract}


\section{INTRODUCTION}
\label{sec:intro}  % \label{} allows reference to this section

Myocardial perfusion single-photon emission computed tomography (MPS) is a widely used modality for evaluating patients with coronary artery disease (CAD)\cite{zaret2010clinical}. The major clinical task in diagnosing CAD using MPS is the detection of myocardial perfusion defects on a single-photon emission computed tomography (SPECT) scan. However, several studies have shown that the diagnostic utility of MPS can be degraded in patients who display certain outlier anatomical characteristics such as large body habitus or large breasts. This is often caused by attenuation artifacts that can adversely impact accurate interpretation \cite{berman2006diagnostic,Burrell193}. Thus, there is an important need to develop methodologies for imaging these patients so that performance on the task of detecting perfusion defects is not degraded.

One approach that could potentially improve this performance is adapting the protocol for image acquisition in a manner that is specific to each patient \cite{4389806,7947224}. In this context, we note that current standard clinical protocols for image acquisition and reconstruction are similar across most patients, with the exception of increased radiotracer dosage or total acquisition time, the former of which increases patient radiation exposure \cite{Henzlova2016}. More specifically, these protocols typically do not account for a patient’s specific anatomy or physiology. In MPS, the clinical protocol involves acquiring data over multiple projection angles, where the gamma camera spends the same amount of time per angle. At each projection angle, the camera detects the number of photons that are emitted by the tracer and reach the detector. Depending on the amount of attenuation of the photons and distance between the camera and the heart, the number and information content of photons detected may be different at each angle. On this topic, Ghaly et al. \cite{Ghaly2411} showed that increasing the amount of time spent on projection angles close to the heart improved defect detection task performance for an ideal observer (IO) \cite{Kupinski:03}. Attenuation in each projection view, in particular, may be different depending on the body habitus. Further, we note that typical MPS procedures involve the acquisition of a low-dose, non-diagnostic CT scan for attenuation compensation \cite{Dorbala2018}. Notably, this CT scan provides a description of the anatomical characteristics of the patient. Hence, the anatomical information provided by this CT scan may present an opportunity to optimize the SPECT acquisition protocol. Based on these ideas, the primary objective of this work is to evaluate if an acquisition protocol optimized based on patient anatomy could lead to improved performance in patients who display outlier anatomical characteristics.

One strategy to perform optimization of acquisition protocols could be the use of an IO with projection data, so that information content in the data for the specific task is maximized \cite{4389806, Kupinski:03, Ghaly_2016}. However, while the use of an IO provides a mechanism to maximize the information content in the projection data for specific tasks, it is unclear whether that translates to improvement for human observers on reconstructed images. As an example, for a signal-known-exactly/background-known-exactly (SKE/BKE) task, an IO may suggest that acquisition at a single projection angle is the most optimal protocol to perform the detection task. This is because the IO can maximize detectability by staying in a single, high-information projection angle throughout the entire acquisition period. However, such a strategy does not lend itself to proper tomographic reconstruction, much less improve the performance of a human observer. Since, in the clinic, the detection task for MPS is performed by human observers on reconstructed images, we designed and implemented a strategy that, while being guided by the IO, still captures data at multiple projection views. We term this strategy to personalize the acquisition protocol as precision SPECT (PRESPECT). PRESPECT integrates the theory of IO with the constraints of tomographic reconstruction. To then test the hypothesis of whether patient-specific acquisition protocols could lead to improved defect detection performance, we conducted a clinically realistic simulation study that compared PRESPECT and standard clinical protocol on the task of detecting perfusion defects in MPS images with an anthropomorphic model observer. Our results provide support to the idea that patient-specific acquisition protocols can improve performance on the defect detection task in MPS.

\section{METHODS}
\label{sec:methods}  % \label{} allows reference to this section

\begin{figure}[t]
    \centering
    \includegraphics[width=1\linewidth]{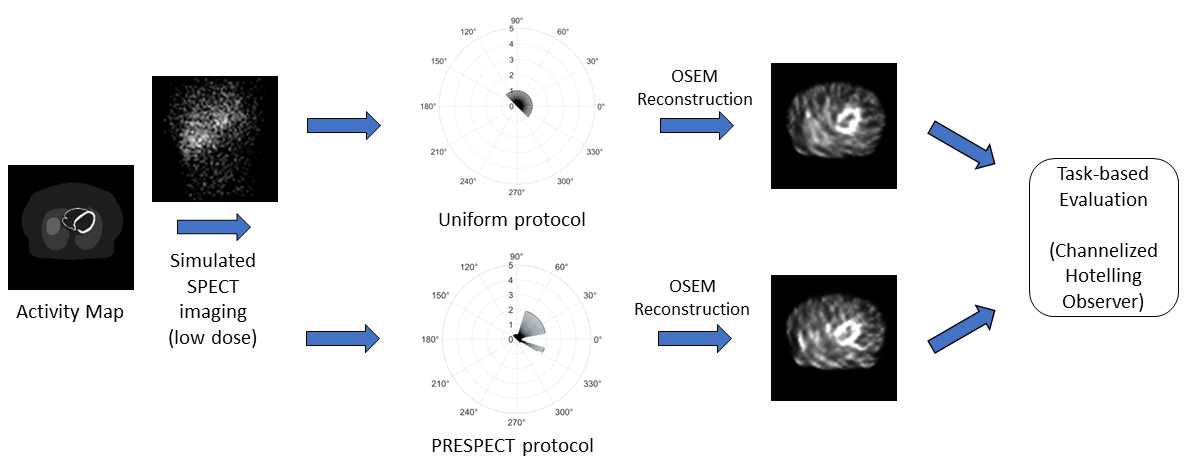}
    \caption{Study design schematic comparing performance of our patient-specific protocol (PRESPECT) and uniform protocol, performed for all five anatomies analyzed.}
    \label{fig:schem}
\end{figure}

We now describe the study conducted to answer our central question of whether a patient-specific acquisition protocol can improve performance on the task of defect detection in MPS as compared to a standard clinical acquisition protocol that spends the same amount of time per projection view. We refer to the standard clinical protocol as uniform protocol. For this, we implemented our proposed patient-specific optimization, namely PRESPECT, and then compared the defect detection task performance of PRESPECT and uniform protocol. Figure \ref{fig:schem} illustrates a schematic of our study design.

\subsection{Study Population}
As we sought to evaluate the impact of a protocol tailored to a patient’s anatomy, we performed our evaluation on a patient-specific basis. We considered a total of five patient anatomies. To obtain our large breast outlier anatomies, we simulated a large population of female patients, considering population distributions for features such as body dimensions, left ventricle size \cite{barclay1995emory,Ghaly_2014}, left ventricle orientation (as angles relative to transverse, sagittal, and coronal planes) \cite{CURTIN201712}, and breast volume \cite{breastvol}. Five patients with large breast volumes greater than 99\textsuperscript{th} percentile were then selected from this population. 

Our goal was to evaluate whether personalizing the protocol would improve defect detection performance for each of these patients. To study this, for each selected outlier patient, we used the 3-D XCAT (extended cardiac and torso) software \cite{XCAT1,XCAT2} to generate multiple physiological realizations that were consistent with the anatomy being considered. Activity in the various organs of the patient was different across different physiological realizations \cite{1282086}. Each generated physiological realization phantom had dimensions 256 × 256 × 256, with a voxel size of 0.17cm × 0.17cm × 0.17cm. These different physiological realizations yielded data to obtain the personalized protocol for each patient anatomy. We also generated a separate patient-specific test set, which consisted of 400 defect-present and 400 defect-absent varying physiological realizations.

\subsection{Proposed PRESPECT Image Acquisition Protocol}
To obtain a personalized acquisition protocol through PRESPECT, we performed the following steps. For a given outlier anatomy, we first considered a specific physiological realization. For that physiological realization, we calculated IO detectability for the defect detection task at each projection angle. This was done using a Fisher information-based approach \cite{Clarkson:10, Jha:13}. Because the specific defect and organ activities would not be known prior to imaging, this IO detectability was computed and averaged over a population of varying defect types and physiological realizations, all of which corresponded to the same patient anatomy. 

From this average IO detectability at each projection view, we performed a constrained optimization to maximize the total IO detectability over all projection views. More specifically, let $N$ denote the total number of projection views, $d_n$, $t_n$ denote the averaged IO detectability and acquisition time at the $n$\textsuperscript{th} projection view, and $T$ denote total acquisition time. Defining $\mathbf{t} = (t_1, t_2, \ldots t_N)$, an $N$-dimensional vector, as the time per projection view, we performed optimization of acquisition time per projection view with the following criteria

\begin{equation}
    \mathbf{t}^\ast = \underset{\mathbf{t}}{\text{argmax}} \; \sum_{n=1}^N t_n d_n \;\;
    \text{s.t.} \; \sum_{n=1}^N t_n = T \text{ and } t_{min} \leq t_n \leq t_{max} \; \forall n.
\end{equation}

Thus, the optimization was conducted under the constraints that the total imaging time stayed the same as in uniform protocol, and also imposing a minimum and maximum scanning time, $t_{min}$ and $t_{max}$, for all projection angles. The latter was done to ensure tomographic reconstruction, as otherwise the protocol may not spend any time on certain projection angles. The specific constraints of minimum and maximum scanning time were chosen for each anatomy through a selection step. 

\subsection{Implementation of Considered Protocols}
For each of the five outlier anatomies, we created a test set of physiological realizations separate from those used for obtaining the average IO detectability. For each test set physiological realization, we simulated the SPECT projection acquisition process using SIMIND, a well-validated Monte Carlo simulation software \cite{LJUNGBERG1989257}. A GE Optima NM/CT 670 scanner, performing SPECT/CT scans, with GE low-energy high-resolution parallel-hole collimators, was simulated. Resulting projections were scaled to 10\% dose level, and Poisson noise was added to generate the final projections. Both PRESPECT and uniform protocol projections were acquired in this manner. We considered a low-dose setting to make the defect detection task challenging, thus providing a setup to assess if PRESPECT yielded improvements in performance. Another reason for choosing a low-dose setting was to investigate the performance of PRESPECT in such settings.

Generated low-count projection data were reconstructed using a 3-D ordered subsets expectation maximization (OSEM) method with attenuation and detector response compensation. For PRESPECT, OSEM was modified to account for the difference in acquisition time per projection angle during the calculation of the forward projection\cite{respgate}. 

\subsection{Image Interpretation and Statistical Analysis}
To assess performance on the defect detection task, we used an anthropomorphic model observer, the channelized Hotelling observer (CHO) with rotationally symmetric channels. Rotationally symmetric channels were chosen because of their emulation of human observer performance on defect detection for MPS \cite{1166633, Sankaran432}, even for the low-count domain at about 10\% of clinical count levels \cite{819288}. These channels have been used in several previous studies \cite{7795176, rahman2023demist,16407,12.2655629,12.2613134}.

For the CHO, we implemented a similar procedure to previous studies that have demonstrated emulation of human observer performance. Reconstructions were reoriented to short-axis orientation and windowed, yielding $32\times32$ regions of the slice containing the centroid of the defect, with the window centered at this defect centroid. Applying the anthropomorphic channels to this region, we obtained test statistics of the CHO for both defect present and absent images, using a leave-one-out training approach. Then, receiver operating characteristic curves (ROC) and area under the ROC curve (AUC) were obtained using the pROC package \cite{Robin2011}. 

The AUC was used as the figure of merit for evaluating the performance of the selected optimized protocol versus uniform protocol for the test set. We used DeLong’s test to assess the differences in AUC values obtained by the two protocols, with a p-value less than 0.01 considered to be statistically significant.

\section{RESULTS}
\label{sec:results}
\begin{table}[b]
\caption{AUC values for uniform vs. PRESPECT protocols on the task of defect detection for the five anatomies considered. 95\textsuperscript{th} percentile confidence intervals are provided in parentheses.}
\label{tab:AUCandDeLongs}
\begin{center}
\begin{adjustbox}{width=0.5\textwidth}
\begin{tabular}{|l|l|l|}
\hline \rule[-1ex]{0pt}{3.5ex}
Anatomy & Uniform AUC & PRESPECT AUC \\ \hline 
\rule[-1ex]{0pt}{3.5ex}
1 & 0.51 (0.49, 0.53) & 0.61 (0.59, 0.63) \\ \hline
\rule[-1ex]{0pt}{3.5ex}
2 & 0.53 (0.51, 0.55) & 0.63 (0.61, 0.65) \\ \hline
\rule[-1ex]{0pt}{3.5ex}
3 & 0.60 (0.58, 0.62) & 0.68 (0.66, 0.70) \\ \hline 
\rule[-1ex]{0pt}{3.5ex}
4 &  0.63 (0.61, 0.65) & 0.73 (0.71, 0.75) \\ \hline
\rule[-1ex]{0pt}{3.5ex}
5 &  0.66 (0.64, 0.68) & 0.75 (0.73, 0.77) \\ \hline 
\end{tabular}
\end{adjustbox}
\end{center}
\end{table}

Table \ref{tab:AUCandDeLongs} shows the AUC values for both protocols. At the considered dose level of 10\%, the performance of PRESPECT significantly outperformed the uniform protocol for all five anatomies, with $p < 0.0001$. The low AUCs overall are likely explained by the challenging nature of large breast anatomies for MPS and the low dose level. Additionally, we observed that PRESPECT improved performance across a range of AUC values, thus providing evidence of patient-specific protocol optimization being able to improve performance at varying levels of difficulty for the defect detection task.

%We also note that the anatomy with the largest myocardium or left ventricle dimensions (anatomy 5) yielded higher AUC for both protocols, and the anatomy with smaller myocardium or left ventricle dimensions (anatomy 3) had lower AUC. Variations in these anatomical dimensions could be a reason for varying AUCs despite all five anatomies being large breast outliers.

\begin{figure}[t]
    \centering
    \includegraphics[width=0.6\linewidth]{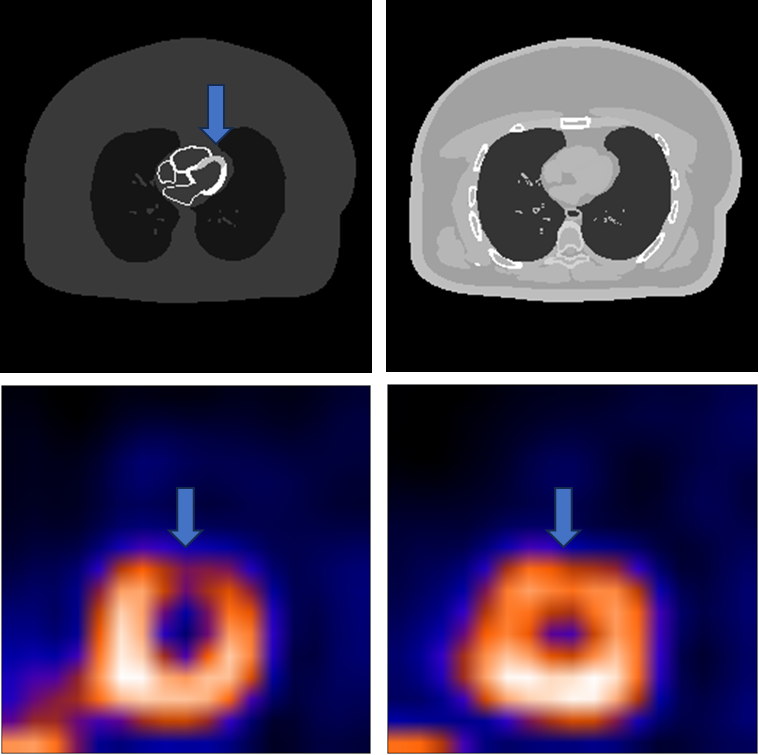}
    \caption{Reconstructed images for both protocols for anatomy 1. A realization of the ground truth activity map with defect indicated (top left), attenuation map (top right), and reconstructed short-axis images, centered at the anterior defect, for PRESPECT (bottom left) and uniform (bottom right) protocols are shown. Acquisition with PRESPECT preserved the defect, which was not present in the uniform protocol reconstruction.}
    \label{fig:asheqrecon}
\end{figure}

We also present sample reconstructions to qualitatively illustrate the improvements made by PRESPECT. Short-axis images for a selected anatomy, with images centered at the defect centroid, are shown in Figure \ref{fig:asheqrecon}. We observed that this example had defect features on PRESPECT acquisition that were not present in the uniform protocol acquisition, further supporting the results observed by the AUC analysis.

\section{CONCLUSIONS AND DISCUSSIONS}
\label{sec:conc}
The results from this study suggest that defect detection performance can be improved by using a personalized acquisition protocol as compared to standard clinical protocol for MPS. We noted that improvements made with PRESPECT were statistically significant and occurred over a range of AUC values. Additionally, assessing sample reconstructions suggested that a personalized protocol optimized through PRESPECT may be able to better preserve the defect signal after acquisition and reconstruction. These findings encourage further exploration of IO-guided optimization to investigate improvements for anthropomorphic observers post-reconstruction, provided suitable constraints are imposed to enable acquisitions at multiple views. Further, while these results were observed for MPS in this study, these findings also encourage similar studies for other imaging modalities. 

In this study, we optimized patient-specific acquisition protocol for a clinically underrepresented group of large breast outliers. The outlier patients we studied do not represent the average female patient, having breast volume above 99\textsuperscript{th} percentile of a normative distribution \cite{breastvol}. Currently, the majority of imaging protocols are developed for the general patient population, and to the best of our knowledge few studies have been conducted on improving medical imaging outcomes specifically focused on patients who may be clinically underrepresented \cite{Bullock-Palmer2022}. Our findings thus motivate research on approaches to design optimal patient-specific acquisition and reconstruction protocols for other clinically underrepresented subgroups.

This early-stage study has several limitations. First, evaluation was done in a low-dose MPS setting. While this study design was chosen to define a clinically challenging task, a study conducted at normal-dose levels would lead to more relevant findings for currently used clinical protocols. This is an area of future investigation. Additionally, results are currently limited to simulated XCAT anatomies, rather than actual patients. Further, the evaluation study was conducted with data from simulated imaging systems and the use of model observers. Conducting this analysis using actual patient images and with human observers would help address these limitations, and would be an important area of additional research. Our results encourage such studies going forward.  

In conclusion, we observed in this study that optimizing acquisition protocol specific to a patient's anatomy can improve defect detection performance for MPS on reconstructed images for patients displaying outlier anatomical characteristics. This provides support for additional investigation into personalized imaging on MPS studies for outlier patients, and motivates additional investigation of similar protocol optimizations for other medical imaging modalities and applications. As with PRESPECT, IO-guided approaches may provide a mechanism for designing such optimization strategies.

\acknowledgments % equivalent to \section*{ACKNOWLEDGMENTS}       

This work was supported by the National Institute of Biomedical Imaging and Bioengineering of the National Institute of Health under grants R01-EB031051 and R01-EB031962, and the NSF CAREER award 2239707. We thank Washington University Research Infrastructure Services for providing computational and storage resources used to conduct this research. The
authors also thank Dr. Paul Segars (paul.segars@duke.edu) and Duke University for providing access to the XCAT phantom software.
% References

\bibliography{report} % bibliography data in report.bib

\begin{thebibliography}{10}

\bibitem{zaret2010clinical}
Zaret, B.~L. and Beller, G.~A.,  [{\em Clinical nuclear cardiology: state of the art and future directions}{\nolinebreak\hspace{0.1em}]}, Elsevier Health Sciences (2010).

\bibitem{berman2006diagnostic}
Berman, D.~S., Kang, X., Nishina, H., Slomka, P.~J., Shaw, L.~J., Hayes, S.~W., Cohen, I., Friedman, J.~D., Gerlach, J., and Germano, G., ``Diagnostic accuracy of gated {Tc-99m} sestamibi stress myocardial perfusion {SPECT} with combined supine and prone acquisitions to detect coronary artery disease in obese and nonobese patients,'' {\em Journal of Nuclear Cardiology}~{\bf 13}(2),  191--201 (2006).

\bibitem{Burrell193}
Burrell, S. and MacDonald, A., ``Artifacts and pitfalls in myocardial perfusion imaging,'' {\em Journal of Nuclear Medicine Technology}~{\bf 34}(4),  193--211 (2006).

\bibitem{4389806}
Barrett, H.~H., Furenlid, L.~R., Freed, M., Hesterman, J.~Y., Kupinski, M.~A., Clarkson, E., and Whitaker, M.~K., ``Adaptive {SPECT},'' {\em IEEE Transactions on Medical Imaging}~{\bf 27}(6),  775--788 (2008).

\bibitem{7947224}
Ghanbari, N., Clarkson, E., Kupinski, M., and Li, X., ``Optimization of an adaptive {SPECT} system with the scanning linear estimator,'' {\em IEEE Transactions on Radiation and Plasma Medical Sciences}~{\bf 1}(5),  435--443 (2017).

\bibitem{Henzlova2016}
Henzlova, M.~J., Duvall, W.~L., Einstein, A.~J., Travin, M.~I., and Verberne, H.~J., ``{ASNC} imaging guidelines for {SPECT} nuclear cardiology procedures: Stress, protocols, and tracers,'' {\em Journal of {Nuclear} {Cardiology}}~{\bf 23},  606--639 (Jun 2016).

\bibitem{Ghaly2411}
Ghaly, M., Links, J., Du, Y., and Frey, E., ``Optimization of {SPECT} using variable acquisition duration,'' {\em Journal of Nuclear Medicine}~{\bf 53}(supplement 1),  2411--2411 (2012).

\bibitem{Kupinski:03}
Kupinski, M.~A., Hoppin, J.~W., Clarkson, E., and Barrett, H.~H., ``Ideal-observer computation in medical imaging with use of {Markov}-chain {Monte} {Carlo} techniques,'' {\em Journal of the Optical Society of America A}~{\bf 20},  430--438 (Mar 2003).

\bibitem{Dorbala2018}
Dorbala, S., Ananthasubramaniam, K., Armstrong, I.~S., Chareonthaitawee, P., DePuey, E.~G., Einstein, A.~J., Gropler, R.~J., Holly, T.~A., Mahmarian, J.~J., Park, M.-A., Polk, D.~M., Russell, R., Slomka, P.~J., Thompson, R.~C., and Wells, R.~G., ``Single photon emission computed tomography ({SPECT}) myocardial perfusion imaging guidelines: Instrumentation, acquisition, processing, and interpretation,'' {\em Journal of {Nuclear} {Cardiology}}~{\bf 25},  1784--1846 (Oct 2018).

\bibitem{Ghaly_2016}
Ghaly, M., Du, Y., Links, J.~M., and Frey, E.~C., ``Collimator optimization in myocardial perfusion {SPECT} using the ideal observer and realistic background variability for lesion detection and joint detection and localization tasks,'' {\em Physics in Medicine \& Biology}~{\bf 61},  2048 (Feb 2016).

\bibitem{barclay1995emory}
Barclay, A., Eisner, R., and DiBella, E., ``Emory {PET} thorax model database,'' (1995).
\newblock Database from Carlyle Fraser Heart Center/Crawford Long Hospital of Emory University and Georgia Institute of Technology.

\bibitem{Ghaly_2014}
Ghaly, M., Du, Y., Fung, G. S.~K., Tsui, B. M.~W., Links, J.~M., and Frey, E., ``Design of a digital phantom population for myocardial perfusion {SPECT} imaging research,'' {\em Physics in Medicine \& Biology}~{\bf 59},  2935 (May 2014).

\bibitem{CURTIN201712}
Curtin, A.~E., Burns, K.~V., Gage, R.~M., and Bank, A.~J., ``Left ventricular orientation and position in an advanced heart failure population,'' {\em Translational {Research} in {Anatomy}}~{\bf 7},  12--19 (2017).

\bibitem{breastvol}
Celeste E.~Coltman, J. R.~S. and McGhee, D.~E., ``Breast volume is affected by body mass index but not age,'' {\em Ergonomics}~{\bf 60}(11),  1576--1585 (2017).
\newblock PMID: 28532249.

\bibitem{XCAT1}
Segars, W.~P., Sturgeon, G., Mendonca, S., Grimes, J., and Tsui, B. M.~W., ``{4D} {XCAT} phantom for multimodality imaging research,'' {\em Medical Physics}~{\bf 37}(9),  4902--4915 (2010).

\bibitem{XCAT2}
Segars, W.~P., Bond, J., Frush, J., Hon, S., Eckersley, C., Williams, C.~H., Feng, J., Tward, D.~J., Ratnanather, J.~T., Miller, M.~I., Frush, D., and Samei, E., ``Population of anatomically variable {4D} {XCAT} adult phantoms for imaging research and optimization,'' {\em Medical Physics}~{\bf 40}(4),  043701 (2013).

\bibitem{1282086}
He, X., Frey, E., Links, J., Gilland, K., Segars, W., and Tsui, B., ``A mathematical observer study for the evaluation and optimization of compensation methods for myocardial {SPECT} using a phantom population that realistically models patient variability,'' {\em IEEE Transactions on Nuclear Science}~{\bf 51}(1),  218--224 (2004).

\bibitem{Clarkson:10}
Clarkson, E. and Shen, F., ``Fisher information and surrogate figures of merit for the task-based assessment of image quality,'' {\em Journal of the Optical Society of America A}~{\bf 27},  2313--2326 (Oct 2010).

\bibitem{Jha:13}
Jha, A.~K., Clarkson, E., and Kupinski, M.~A., ``An ideal-observer framework to investigate signal detectability in diffuse optical imaging,'' {\em Biomedical Optics Express}~{\bf 4},  2107--2123 (Oct 2013).

\bibitem{LJUNGBERG1989257}
Ljungberg, M. and Strand, S.-E., ``A {Monte} {Carlo} program for the simulation of scintillation camera characteristics,'' {\em Computer Methods and Programs in Biomedicine}~{\bf 29}(4),  257--272 (1989).

\bibitem{respgate}
Qi, W., Yang, Y., Wernick, M.~N., Pretorius, P.~H., and King, M.~A., ``Limited-angle effect compensation for respiratory binned cardiac {SPECT},'' {\em Medical Physics}~{\bf 43}(1),  443--454 (2016).

\bibitem{1166633}
Frey, E., Gilland, K., and Tsui, B., ``Application of task-based measures of image quality to optimization and evaluation of three-dimensional reconstruction-based compensation methods in myocardial perfusion {SPECT},'' {\em IEEE Transactions on Medical Imaging}~{\bf 21}(9),  1040--1050 (2002).

\bibitem{Sankaran432}
Sankaran, S., Frey, E.~C., Gilland, K.~L., and Tsui, B.~M., ``Optimum compensation method and filter cutoff frequency in myocardial {SPECT}: A human observer study,'' {\em Journal of {Nuclear} {Medicine}}~{\bf 43}(3),  432--438 (2002).

\bibitem{819288}
Wollenweber, S., Tsui, B., Lalush, D., Frey, E., LaCroix, K., and Gullberg, G., ``Comparison of {Hotelling} observer models and human observers in defect detection from myocardial {SPECT} imaging,'' {\em IEEE Transactions on Nuclear Science}~{\bf 46}(6),  2098--2103 (1999).

\bibitem{7795176}
Li, X., Jha, A.~K., Ghaly, M., Elshahaby, F. E.~A., Links, J.~M., and Frey, E.~C., ``Use of sub-ensembles and multi-template observers to evaluate detection task performance for data that are not multivariate normal,'' {\em IEEE Transactions on Medical Imaging}~{\bf 36}(4),  917--929 (2017).

\bibitem{rahman2023demist}
Rahman, M.~A., Yu, Z., Laforest, R., Abbey, C.~K., Siegel, B.~A., and Jha, A.~K., ``{DEMIST}: A deep-learning-based task-specific denoising approach for myocardial perfusion {SPECT},'' (2023).
\newblock arXiv:2306.04249 [physics.med-ph].

\bibitem{16407}
Yu, Z., Rahman, M.~A., Laforest, R., Schindler, T.~H., Gropler, R.~J., Wahl, R.~L., Siegel, B.~A., and Jha, A.~K., ``Need for objective task-based evaluation of deep learning-based denoising methods: A study in the context of myocardial perfusion {SPECT},'' {\em Medical Physics}~{\bf 50}(7),  4122--4137 (2023).

\bibitem{12.2655629}
Rahman, M.~A., Yu, Z., Siegel, B.~A., and Jha, A.~K., ``{A task-specific deep-learning-based denoising approach for myocardial perfusion SPECT},'' in [{\em Medical Imaging 2023: Image Perception, Observer Performance, and Technology Assessment}{\nolinebreak\hspace{0.1em}]},  Mello-Thoms, C.~R. and Chen, Y., eds.,  {\bf 12467},  1246719, International Society for Optics and Photonics, SPIE (2023).

\bibitem{12.2613134}
Yu, Z., Rahman, M.~A., and Jha, A.~K., ``{Investigating the limited performance of a deep-learning-based SPECT denoising approach: an observer-study-based characterization},'' in [{\em Medical Imaging 2022: Image Perception, Observer Performance, and Technology Assessment}{\nolinebreak\hspace{0.1em}]},  Mello-Thoms, C.~R. and Taylor-Phillips, S., eds.,  {\bf 12035},  120350D, International Society for Optics and Photonics, SPIE (2022).

\bibitem{Robin2011}
Robin, X., Turck, N., Hainard, A., Tiberti, N., Lisacek, F., Sanchez, J.-C., and M{\"u}ller, M., ``{pROC}: an open-source package for {R} and {S}+ to analyze and compare {ROC} curves,'' {\em BMC Bioinformatics}~{\bf 12},  77 (Mar 2011).

\bibitem{Bullock-Palmer2022}
Bullock-Palmer, R.~P., Peix, A., and Aggarwal, N.~R., ``Nuclear cardiology in women and underrepresented minority populations,'' {\em Current Cardiology Reports}~{\bf 24},  553--566 (May 2022).

\end{thebibliography}
\bibliographystyle{spiebib} % makes bibtex use spiebib.bst

\end{document}